\documentclass[aps,pre,twocolumn,showpacs,superscriptaddress]{revtex4}
\usepackage[]{graphicx}
\begin{document}

\title{Prediction of high frequency intrinsic localized modes in Ni and Nb}

\author{M. Haas$^1$, V. Hizhnyakov$^1$, A. Shelkan}
\affiliation{Institute of Physics, University of Tartu, Riia 142,
51014 Tartu, Estonia}
\author{M. Klopov}
\affiliation{Institute of Physics, Tallinn University of Technology,
Ehitajate 5, 19086 Tallinn, Estonia}
\author{A. J. Sievers}
\affiliation{Laboratory of Atomic and Solid State Physics, Cornell University, USA}

\date{\today}

\begin{abstract}

It is found that in some metals an intrinsic localized mode may exist with frequency
above the top of the phonon spectrum. The necessary condition, requiring
sufficiently high ratio of quartic to cubic anharmonicity may be fulfilled
because of screening of the interaction between ions by free electrons.
Starting from the known literature values of the pair potentials we have
found that in Ni and Nb the derived localized mode condition is fulfilled.
MD simulations of the nonlinear dynamics of Ni and Nb confirmed that high frequency
ILMs may exist in these metals.

\end{abstract}


\maketitle

\draft

\section{Introduction}
The study of vibrational energy localization in highly excited small
molecules is well known and a number of reviews have appeared [1-4]. The
idea of localization in perfect anharmonic lattices was considered by
Kosevich and Kovalev [5] for the case of a monatomic chain with nearest
neighbor harmonic, cubic and quartic anharmonic interactions. They showed that
an envelop soliton-like excitation with frequency above the top of the
phonon band could exist for sufficiently weak cubic interactions. Somewhat
later strongly localized vibrational modes in anharmonic lattices were
proposed in Refs. [6,7]. The realization that this new excitation phenomenon
only required nonlinearity plus discreteness expanded the topic in different
directions, ranging from analytical considerations [6-9] to MD simulations
[10-12]. The early reviews of the resulting field focused on predicting
different processes [13-15]. These excitations, which are often referred to
as intrinsic localized modes (ILMs), discrete breathers or discrete
solitons, have now been identified in different driven physical systems
including electronic and magnetic solids, Josephson junctions,
micromechanical arrays, optical waveguide arrays, and laser-induced photonic
crystals [16-18].

In numerical studies of ILMs in atomic lattices different two-body potential
models such as Lenard-Jones, Born-Mayer-Coulomb, Toda, and Morse potentials
as well as their combinations have been used in the past. All of these
potentials show strong softening with increasing vibrational amplitude and
the ILMs, found in these simulations, always drop down from
the optical band(s) into the phonon gap, if there is one. (See Refs.
[19-22], where ILMs in alkali halide crystals have been calculated).
Consequently, it has been assumed that the softening of atomic bonds with
increasing vibrational amplitude is a general property of crystals and
therefore ILMs with frequencies above the top phonon frequency cannot occur.
However recent inelastic neutron scattering investigation of the vibrational
excitations in uranium ($\alpha $-U) in thermal equilibrium showed
some degree of localization near the top of the phonon spectrum at elevated
temperatures [23]. For this to occur the pair potentials in this metal must
be fundamentally different from those describing alkali halide crystals.
Because the electrons at the Fermi surface provide an essential contribution
to the screening of the ion-ion interaction in metals there is no apriori
reason to expect the anharmonicities of these two very different systems to
be similar.

The purpose of this paper is to explore the ILM properties in nickel and
niobium by starting from the embedded atom model (EAM). This technique
allows one to find the potential energy of vibrations in metals, taking the
screening effects into account [24,25]. Our findings show that ILMs may be
expected to occur in both of these metallic crystals. In the next section we
illustrate how the elastic springs of the atomic bonds are to be
renormalized to give the minimal condition for localized mode production. In the
small ILM amplitude limit two additive contributions to the renormalization
are identified: the positive one given by the quartic anharmonicity and the
negative one determined by the square of cubic anharmonicity. The ILM
appears above the phonon spectrum when the first contribution exceeds the
second one. For the monatomic chain with nearest neighbor interactions this condition 
agrees with that found by Kosevich and Kovalev [5] and it corresponds to a rather high 
ratio of quartic to cubic anharmonicities but in 3D lattices a smaller ratio is required. 
Our calculations demonstrate that in Ni and Nb this ILM condition is fulfilled. The
molecular dynamics simulations for nickel and for niobium confirming this result
are described in Section 3 followed by some conclusions in Section 4.

\section{Threshold condition for ILM}
One can readily present an argument in favor of the localized mode possibility
in metals. The point is that the essential contribution to the screening of
the atomic interactions in metals comes from free electrons at the Fermi
surface. Due to their well-defined energy and the oscillating character of
the wave functions of these electrons (Friedel oscillations) the resulting
pair-potentials may acquire non-monotonic, or even oscillatory dependence on
the atomic distance [26]. One consequence is that the ion-ion attractive
force, at intermediate distances, may be enhanced resulting in an
amplification of even anharmonicities for the resulting two-body potentials.
This effect can counteract the underlying softening associated with the bare
potentials with moderate increase of vibrational amplitudes to permit the
existence of ILMs above the top of the phonon spectrum.

Let the anharmonic potential describing the nearest neighbor interactions for a 1-D 
monatomic chain be represented by
\begin{equation}
\label{eq1}
U = \sum_n \sum_{p=2}^{4} \frac{K_p}{p} (u_{n+1}-u_n)^p
\end{equation}
where $K_2$  is the harmonic force constant and $K_3$, $K_4$  are the anharmonic ones,
$u_n$ is the displacement of the $n$-th atom from its equilibrium position.
The condition for the formation of an ILM with the frequency above the top 
of the phonon spectrum, given in Ref. [5], is  $\kappa=3 K_2 K_4/4 K_3^2>1$.

To obtain the minimal condition more generally we use the equation for
renormalization of the elastic springs of the atomic bonds by the ILM derived
in Refs. [27, 28], which is
\begin{equation}
\label{eq2}
\delta K_{2n} =2 \langle \sin ^2\left( {\omega _L t} \right)\,\partial
^2 V^{anh} / \partial r_n^2 \rangle ,
\end{equation}
where $\delta K_{2n} $ is the change produced in the harmonic spring of the pair potential 
of bond number $n$. Here $\omega _L $ is the frequency of ILM, $V^{anh}$ is the anharmonic 
part of the potential, the second derivative is taken for the distance of the bond 
$r_n=r_{0n} +\bar {A}_n \cos (\omega _L t)+\bar {\xi }_n $, where $r_{0n}$ 
is the length of the bond, $\bar {A}_n $ is the amplitude of vibration of this bond,
\begin{equation}
\label{eq3}
\bar {\xi }_n =\sum\limits_{n_l } {\bar {g}_{nn_l } \left\langle {{\partial
V^{anh}} \mathord{\left/ {\vphantom {{\partial V}
{\partial r_{n_l}}}} \right. \kern-\nulldelimiterspace} {\partial r_{n_l}}}
\right\rangle }
\end{equation}
is the dc change (usually extension) of its length due to the ILM,
$\bar
{g}_{nn_l } = g_{nn_l} -g_{n'n_l} -g_{nn'_l} +g_{n'n'_l}$, $n$ and $n'$ are
the indexes of two ends of the bond $n$, $g_{nn_l } =-\left( {M_{n} M_{n_l } }
\right)^{{-1} \mathord{\left/ {\vphantom {{-1} 2}} \right.
\kern-\nulldelimiterspace} 2}\sum\nolimits_i {e_{ni} e_{n_l i} } \omega
_i^{-2} $ is the static limit of the lattice Green's function, $e_{ni} $ is
the polarization vector of the phonon $i$ for the bond $n$, $\omega _i$ is the
frequency of the phonon, $M_n$ is the mass of the atom $n$.

For a small amplitude ILM one need only consider cubic and quartic anharmonicity. In this 
approximation  Eqs. (\ref{eq2}) and (\ref{eq3})
take the form
\begin{equation}
\label{eq4}
\delta K_{2n} =2K_3 \bar {\xi }_n +\frac{3}{4}K_4 \bar {A}_n^2 ,
\end{equation}
\begin{equation}
\label{eq5}
\bar {\xi }_n =\frac{1}{2}\sum\limits_{n_l } {\bar {g}_{nn_l } } K_{3n_l }
\bar {A}_{n_l }^2 .
\end{equation}
Usually the first term in Eq. (4) is negative while the second term is positive. 
The bond will harden with increasing amplitude of vibrations and an ILM will shift up 
from the phonon band if the absolute value of the first term is smaller than the second 
term. To fulfill this condition the value of the parameter $\kappa$ needs to be 
sufficiently large.  This value depends not only on the pair potentials but also 
on the type and dimension of the lattice.

We consider first $\delta K_2 $ in monatomic chain with nearest neighbor interactions. 
In this case $n'=n + 1$ and $g_{nn_l } $ depends on $\left| {n-n_l} \right|$.
Therefore $\bar {g}_{nn_l } =2g_{nn_l } -g_{n+1n_l } -g_{n-1n_l}$. From the equations 
of motion one gets  $\omega _i^2 e_{in} =\left( {{K_2 }
\mathord{\left/ {\vphantom {{K_2 } M}} \right. \kern-\nulldelimiterspace} M}
\right)\left( {2e_{in} -e_{in+1} -e_{in-1} } \right)$. Multiplying both
sides of this equation by $\omega _i^{-2} e_{in_l} $ and summing up over $i$ we
get $\delta _{nn_l } =-K_2 \bar {g}_{nn_l } $. Consequently the dc lattice
expansion equals
\begin{equation}
\label{eq6}
\bar {\xi }=-\left( {{K_3 } \mathord{\left/ {\vphantom {{K_3 } {2K_2 }}}
\right. \kern-\nulldelimiterspace} {2K_2 }} \right)\bar {A}^2
\end{equation}
(index $n$ is now omitted). Therefore
\begin{equation}
\label{eq7}
\delta K_2 =\left( {{3K_4 } \mathord{\left/ {\vphantom {{3K_4 } {4-{K_3^2 }
\mathord{\left/ {\vphantom {{K_3^2 } {K_2 }}} \right.
\kern-\nulldelimiterspace} {K_2 }}}} \right. \kern-\nulldelimiterspace}
{4-{K_3^2 } \mathord{\left/ {\vphantom {{K_3^2 } {K_2 }}} \right.
\kern-\nulldelimiterspace} {K_2 }}} \right)\bar {A}^2.
\end{equation}
The bond will harden with increasing amplitude of vibrations and an ILM will
shift up from the phonon band if $\kappa ={3K_2 K_4 } \mathord{\left/
{\vphantom {{3K_2 K_4 } {4K_3^2 }}} \right. \kern-\nulldelimiterspace}
{4K_3^2 }>1$.  This condition is identical to that found in Ref. [5] for localized 
vibrations in the chain. A detailed discussion of this condition also has been given in 
Ref. [29]. (For arbitrary amplitude the ILM condition for the potential described by 
Eq. (1) is given also in Ref. [30].)

Let us apply now Eqs. (\ref{eq4}) and (\ref{eq5}) to 3-D lattices. We treat an even 
symmetry ILM in a monatomic fcc or bcc lattice with the main motion directed along 
the shortest bond. Note that the high-energy edge of the phonon DOS in both these lattices 
corresponds to the short-wavelength phonons. Therefore the initial ILM under 
consideration, shifting up from the phonon band, resembles a wave packet of the standing 
longitudinal plane waves; it has a large spatial extent.

We assume that the anharmonic interactions are well localized. This allows one to include 
in Eq. (5) only contributions of the shortest bonds in the  $xy$-direction (fcc lattice) 
or in  $xyz$-direction (bcc lattice). The factors  $\bar {g}_{nn_l}$ in this equation 
tend to zero with increasing $\left| {n-n_l} \right|$. Therefore if the ILM is close to 
the threshold limit then the corresponding amplitudes of vibrations of all these bonds 
at the contributing $n_{l}$ sites in Eq. (\ref{eq5}) are almost the same and so 
the dc distortion
\begin{equation}
\label{eq8}
\bar {\xi }\simeq - \frac {2K_3\bar {A}^2}{MN} \sum\limits_q {\sum\limits_{n=0}^{N-1}} \ 
\frac{1-\cos(qr_0)}{\omega_q^2} \, e\,^{iqr_0 (n-N/2)},
\end{equation}
where $Nr_0 \,$ is the distance between the border atoms in the direction of the main 
vibration ($r_0 \,$the equilibrium first-neighbor distance), $\omega _q $ is the 
frequency of the longitudinal waves, $q$ is the wave number acquiring the discrete values 
$q=\pi k/r_0 N$ with $k=-N/2,-N/2+1,\,\,...\,N/2-1$ ($N$ is even). 
In the $N\to \infty $ limit only the term $k=0$ contributes so
\begin{equation}
\label{eq9}
\bar {\xi }=-\left( {{K_3 } \mathord{\left/ {\vphantom {{K_3 } {2\tilde K_2 }}}
\right. \kern-\nulldelimiterspace} {2\tilde K_2 }} \right)\bar {A}^2 ,
\end{equation}
where $\tilde {K}_2 ={Mv_l^2} \mathord{\left/ {\vphantom {{Mv^2} {r_0^2 }}}
\right. \kern-\nulldelimiterspace} {r_0^2 }$ is the mean elastic spring in
the bulk, $v_l$ is the longitudinal velocity of sound. Comparing this equation
for $\bar {\xi }$ with Eq. (\ref{eq6}) and considering that $\tilde {K}_2 $ is
larger than $K_2 $ we conclude that the expansion by an ILM in a 3D
lattice is hindered as compared to the chain, - a physically intuitive
result. Inserting Eq. (\ref{eq9}) for $\bar {\xi }$ into Eq. (\ref{eq4}) we get 
$\delta K_2$. The hardening of the bonds in 3D lattices takes place and an ILM shifts up 
from the phonon band if
\begin{equation}
\label{eq10}
{\tilde {\kappa }= 3 \tilde {K}_2 K_4} \mathord{\left/ {\vphantom {{\tilde {\kappa
}= 3 \tilde {K}_2 K_4} {4K_3^2 }}} \right. \kern-\nulldelimiterspace} {4K_3^2
}>1.
\end{equation}
This condition is easier to fulfill than the 1D condition for $\kappa > 1$.

\section{MD simulations}
\subsection{ILMs in nickel}

The potential energy of Ni developed in Ref. [31] has the customary form 
for the embedded atom model [24, 25]
\begin{equation}
\label{eq11}
E_{tot} =\frac{1}{2}\sum\limits_{n{n}'} {V\left( {r_{n{n}'} } \right)}
+\sum\limits_n {F_n\left( {\bar {\rho}_n} \right)},
\end{equation}
where $V\left( {r_{n{n}'} } \right)\,$ is a pair potential as a function of
the distance $r_{n{n}'} \,$ between atoms $n$ and ${n}'$, $F_n$ is the embedding
energy of atom $n$ as a function of the electron density $\bar{\rho}_n
=\sum\nolimits_{{n}'\ne n} {\rho \left( {r_{n{n}'} } \right)} $ induced at
atom $n$ by all other atoms in the system,
$\rho \left( {r_{n{n}'} } \right)$ is the electron density at atom $n$ due to atom $n'$
as a function of the distance between them.
The second term in 
Eq. (\ref{eq11}) is volume dependent.
\begin{figure}[th]
\includegraphics[angle=0,width=.48\textwidth]{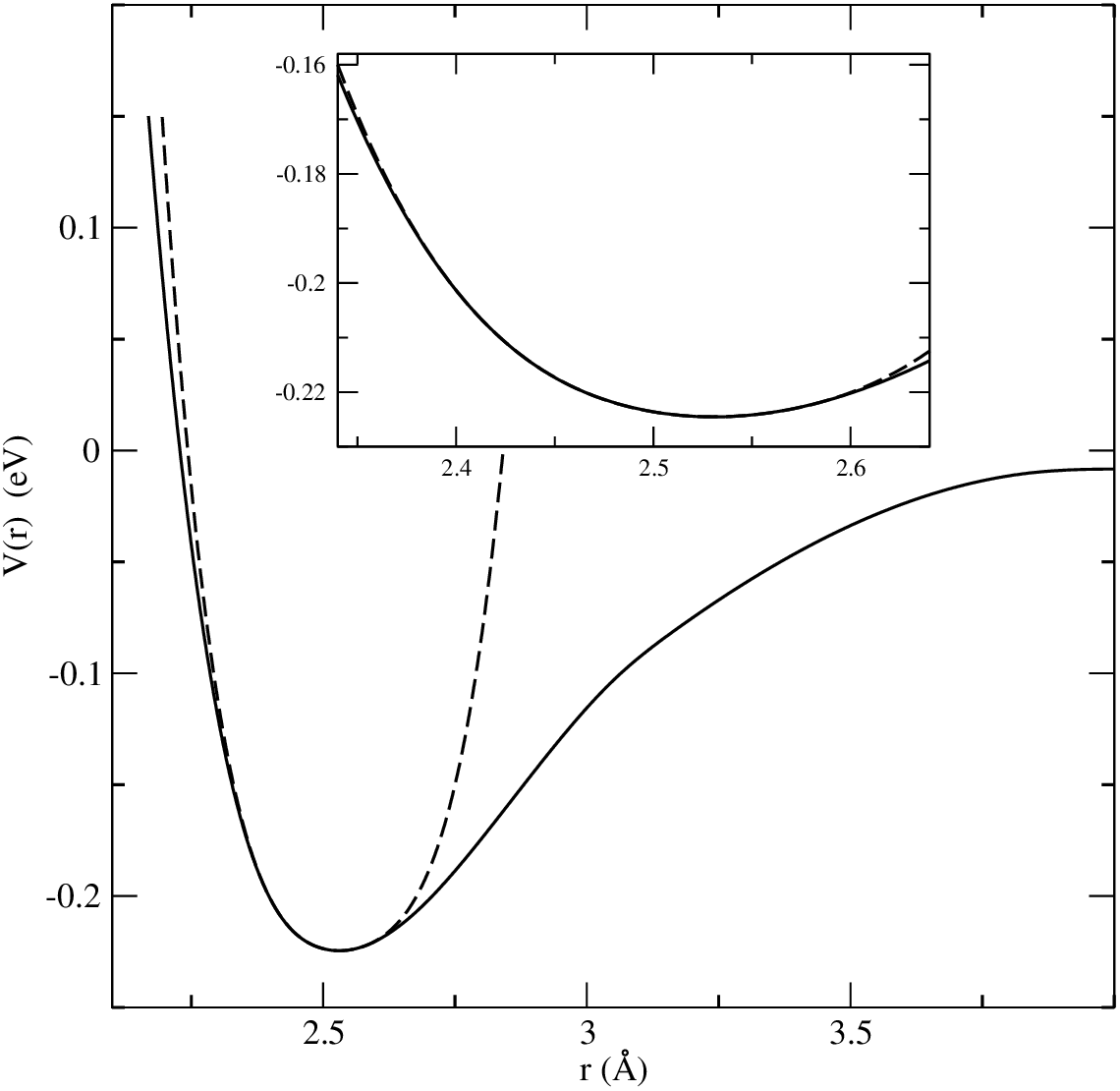}\hspace*{0em}
\caption{The pair potential $V(r)$ of Ni (solid line) and its approximation by the
fourth-order polynomial (dashed line). Insert shows an expanded view.}
\end{figure}
Its contribution is essential for determining the equilibrium
configuration of the lattice. Below we use the  effective pair format in which this term 
has only quadratic and higher-order contributions with respect to the displacement of 
atoms from their equilibrium position in the lattice. For small displacements compared 
to the lattice constant this term is usually small. In Ni corresponding correction of
the elastic springs is of the order of $1\% $, and the
correction of the anharmonic forces is even smaller.

The pair potential $V(r)$ of Ni found from the data placed on the website [32] using the
cubic spline approximation is presented in Fig. 1. In the same figure the approximation
by the fourth-order polynomial using the least-square method in the interval
$2.496 \, \textrm{\AA} \pm 0.116\, \textrm{\AA}$ is also presented. This interval 
corresponds to the actual values of coordinates of the ILM with the frequency near 
the top of the phonon band (see Table I). In this approximation 
$K_2 \approx 2.32 \, \textrm{eV} / {\textrm{\AA}}^2$, 
$K_3 \approx -11 \, \textrm{eV} / {\textrm{\AA}}^3$, 
$K_4 \approx 70 \, \textrm{eV} / {\textrm{\AA}}^4$. 
The root-mean-square deviation for the
polynomial approximation in the given interval is $10^{-5}$ eV, i.e. $\sim$ 1000 times
less than the corresponding vibrational energy of the bond.

The distance between the nearest atoms in Ni at room temperature $r_0 = 2.49 \, 
\textrm{\AA}$ and longitudinal sound velocity $v_l = 5266$ m/sec. These values give 
$\tilde {K}_2 =2.75 \, \textrm{eV} / {\textrm{\AA}}^2$, (as expected 
$\tilde {K}_2 > K_2$) and $\tilde {\kappa }\approx 1.2$. The distance  $r_0$ increases 
with temperature ($r_0 = 2.51 \, \textrm{\AA}$ at $T = 800$ K) while $v_l$ decreases with 
temperature ($v_l=5100$ m/sec at $T = 800$ K). The $\tilde {\kappa }$ value also 
decreases with temperature remaining at $T = 800$ K  somewhat larger than 1. Hence, in Ni 
the conidition $\tilde {\kappa }>1$ is satisfied both at room and at high temperatures. 

A fortunate circumstance is that the phonon DOS in Ni (and in other monatomic fcc 
lattices) has quite a sharp high frequency peak corresponding to the short-wave phonons 
(see Fig. 2) resulting in a straightforward  localization of the wave packet of these 
phonons. Consequently one can expect that in Ni ILMs can exist with the frequency 
above the top of the phonon spectrum and that their amplitudes and hence 
corresponding energies may be relatively small.

\begin{figure}[th]
\includegraphics[angle=0,width=.48\textwidth]{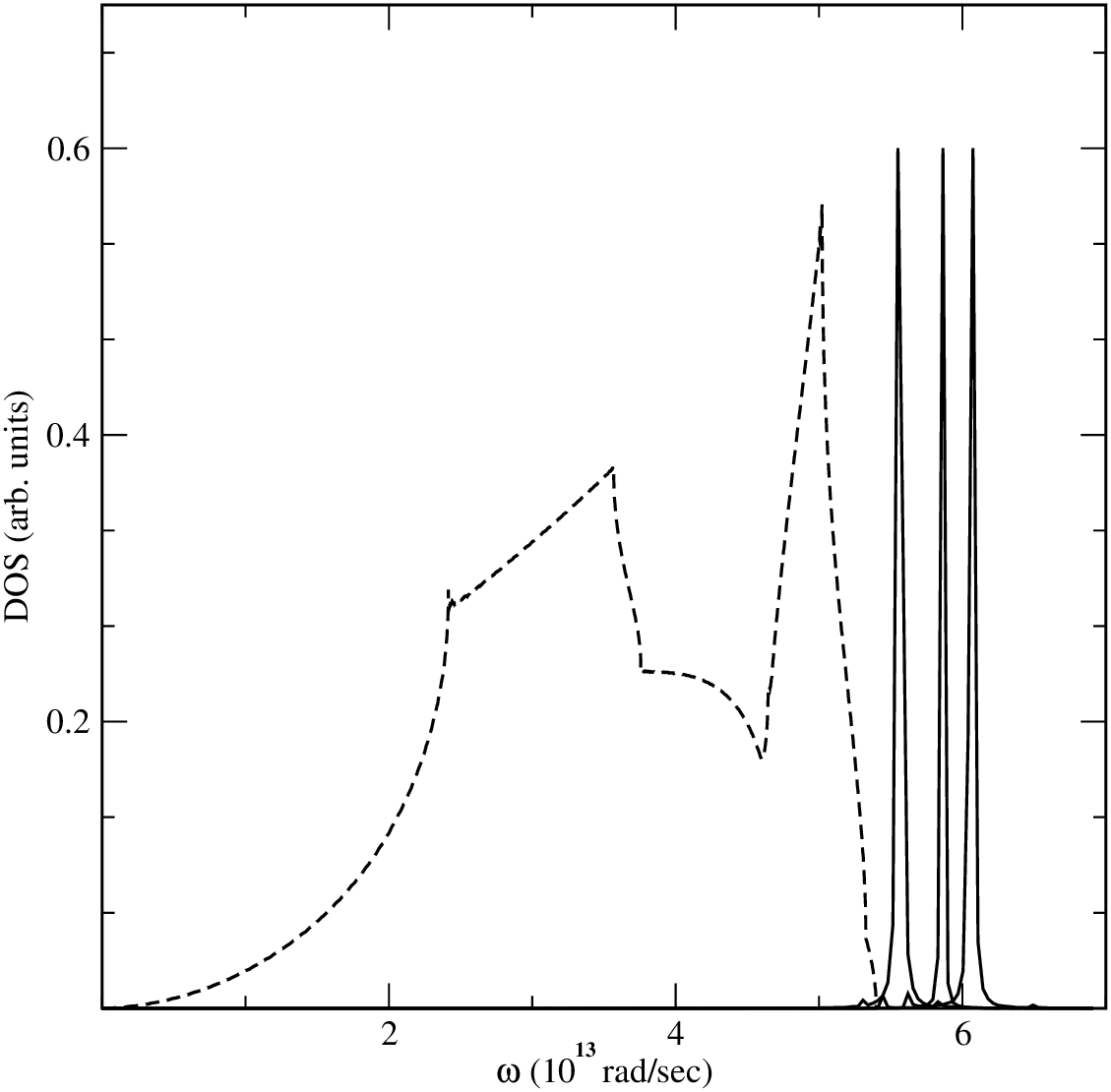}\hspace*{0em}
\caption{Phonon density of states and three ILM spectral signatures for Ni.
Phonon spectrum (dashed line) and spectrographs
(solid line) of the different ILMs:  The frequencies are 5.58, 5.86, 6.07 
(10$^{13}$ rad/sec) and the  amplitudes of vibrations of the central bond are  
0.18 \AA, 0.31 \AA \ and 0.42 \AA.}
\end{figure}
\begin{figure}[th]
\includegraphics[angle=0,width=.48\textwidth]{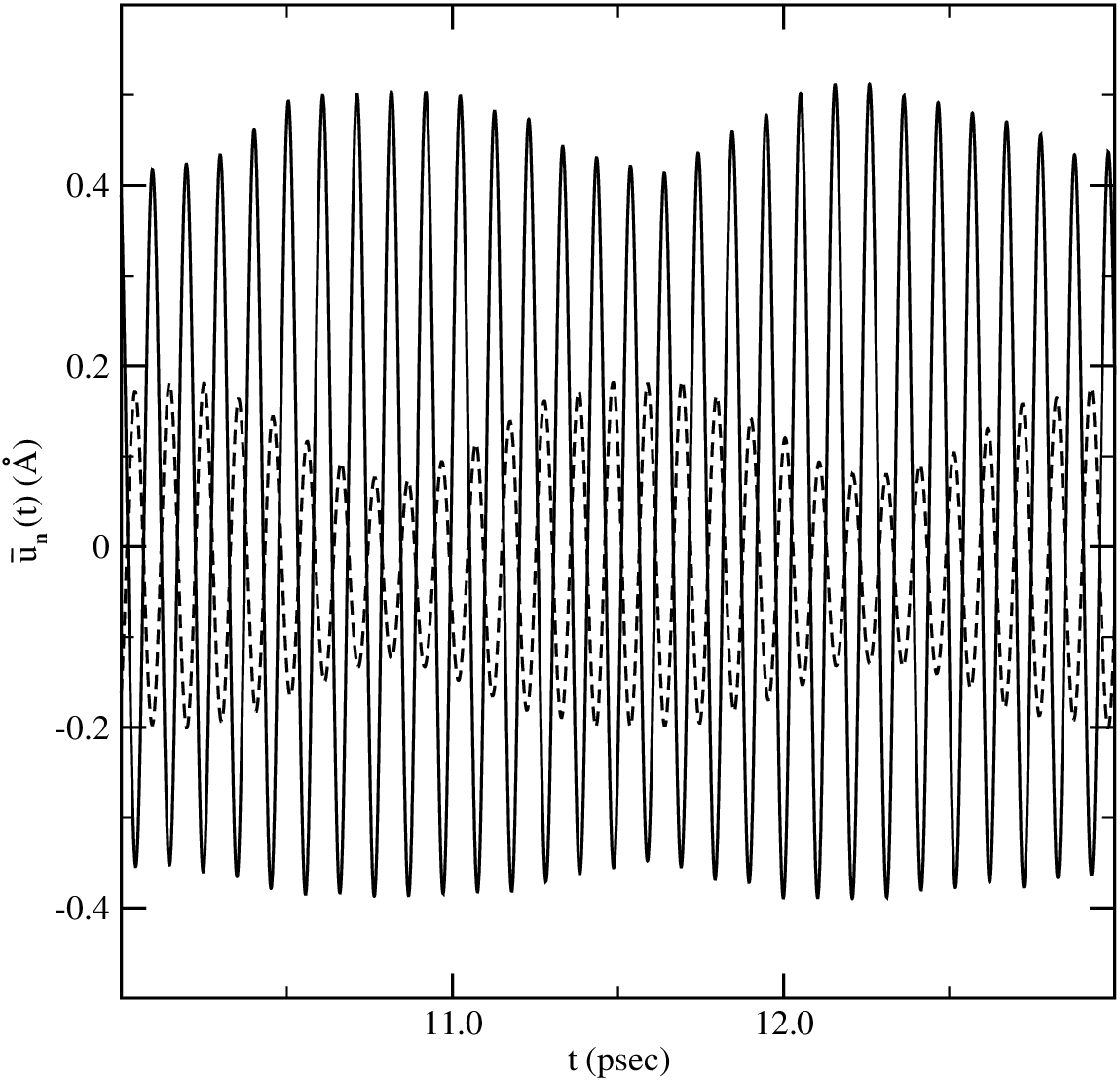}\hspace*{0em}
\caption{Time dependence $\bar {u}_n (t)=r_n(t)-r_0$ of the vibration of the central
($n$=0, solid line) and third ($n$=3, dashed line) bonds in Ni at long times
containing the ILM with the frequency 6.07 $\cdot$ 10$^{13}$ rad/sec.
The corresponding values of the dc distortion $\bar{\xi}_n$ are given in Table I.
The amplitude modulation of the ILM is induced by its partner,
the nearby linear local mode.}
\end{figure}

To verify this prediction we performed MD simulations of vibrations of Ni
clusters using the full two-body potential without polynomial approximation. 
Since the long-range interactions in metals are screened out the
cluster calculations should give reliable results assuming that the size of
the cluster is sufficiently large. In our calculations we studied clusters
up to 22056 atoms with different boundary conditions: a) periodic, b) free
ends, c) fixed ends. The results of these calculations agree well with each
other. Although the second (volume dependent) term in Eq. (\ref{eq11}) gives only
small corrections to forces, it was included in our MD simulations of clusters with 
periodic boundary conditions. 

For the boundary conditions with free and fixed ends only the linear part of the second 
term in Eq. (\ref{eq11}) was considered. We have found that this approximation does not 
noticeably change the results but allows one to significantly shorten the calculation
time. The resulting phonon dispersion curves are in satisfactory agreement with those 
in the literature [31].

Classical molecular dynamics of a Ni cluster was calculated by means of the basic 
Verlet algorithm:
\begin{equation}
\label{eq12}
u(t+dt)=u(t)+v(t)dt+\frac{1}{2}a(t)dt^2,
\end{equation}
\begin{equation}
\label{eq13}
v(t+dt)=v(t)+\frac{1}{2}\left( {a(t)+a(t+dt)} \right)dt.
\end{equation}
Here $t$ is time, $u(t)$ is displacement of the atom from its equilibrium
position, $v(t)$ and $a(t)$ are the velocity and acceleration of the atom.
The latter were found from Newton's second law by calculating the gradients
of $E_{tot} $ with respect to the atom coordinates. For the case of periodic
boundary conditions the periodicity was a cube with edge 52.8 \AA, which
includes 13500 atoms. A time-step 2 fs was used; 7000 time steps
have been calculated which corresponds approximately to 200 periods of
vibrations of the ILM. The results of the calculations are given in Figs. 2 and 3
and in Table I. For the case of free ends the calculated cluster had 34 parallel
$40r_0 \times 17r_0 $ square pallets; it includes 23120 atoms. A time-step
of 0.01 fs was used. A few million time steps were calculated, so that the
full calculated time would contain several hundreds of periods of ILM.

In the first runs to excite the lattice vibrations, 8 nearest central atoms located at 
$\sqrt{2} r_0[n/2, n/2, 0],n=-4,-3,\dots 3$
(in the central chain) have been initially displaced from their equilibrium
position constituting an even structure; the displacements $u_{n}$ of the atoms
from their equilibrium position have been chosen as
follows: $u_0 =-u_1 =u_2 =-2u_3 $. This displacement pattern has the correct symmetry
of the ILM but not the exact shape. The values of $u_0 $ varied from 0.09 \AA \ 
to 0.3 \AA. After the shake off of the phonons at short times
the ILM was recognized as undamped periodic motion at large
times (slowly modulated) at a frequency above the maximum of the
phonon band. As an example of the corresponding time dependence of the
vibration at long times, see Fig. 3, where the calculated dependence on time of the
difference $\bar {u}_n (t)=r_n (t)-r_0$ for two bonds $n$ = 0
and $n$ = 3 is given. 

The dependence of the vibration frequency on amplitude
of the central bonds is presented in Fig. 4; the vibrational amplitudes $\bar
{A}_n $ and the dc expansion $\bar {\xi }_n $ of several bonds in the (110)
direction are given on Table I. We also calculated several bonds in other
directions; the values of corresponding $\bar {A}_n $ and $\bar {\xi }_n $
appeared to be somewhat smaller.
As follows from the Table I all seven  ILMs presented are quite broad spatially,
with the first three essentially keeping the same localization, and the other four ILMs
broadening. Obviously this is a result of the softening of the pair potential as
the amplitude of the ILMs increases. Still the amplitudes of all such ILMs decrease 
rapidly in the  periphery as seen from the $\bar{A}_6$ column.

To control the numerical procedure, the full energy of the cluster was
calculated at every time step. The energy conservation law was found to be
well fulfilled for the entire calculated time interval. We have also
calculated the energy of the ILM; the contribution of 215 atoms situated
in a prolate-ellipsoid shape was included in calculations. It appears that
$\sim $2/3 of the energy of the initially displaced 8 central atoms remains
localized, while again 2/3 of it belongs to the atoms in the
initially excited chain and $\sim $ 1/3 goes to the
nearest surrounding atoms. We have found that an ILM in Ni may have a rather
small energy $\sim $ 0.2 eV (see the first line in the Table I). 
The reason is the existence of the narrow peak in the phonon
DOS belonging to short-wave phonons, which permit a rapid splitting of the
ILM away from the phonon band. On the other hand, for large 
$\bar{A}_0 \geq 0.314 \, \textrm{\AA}$ 
one observes a significant difference between the ILM frequency and the top phonon 
frequency; thus we conclude that the existence of ILMs in Ni is reliable.

\begin{widetext}

Table I. Spatial properties of ILMs in nickel. The difference of the
frequency $\omega _L$ of the even ILM and the maximum phonon
frequency $\omega _M=5.4\cdotp10^{13}$ rad/sec, amplitudes of the bonds $\bar {A}_n$ 
and the changes of their length $\bar {\xi }_n$ for the atoms located at 
$\sqrt{2}r_0[n/2, n/2, 0]$, with $n =0, 1, 2, 3, 6$, and the resulting ILM energy $E$.
The shifts of atoms of the central chain satisfy the condition $u_{-n}=-u_{n-1}$.

\begin{table}[htbp]
\begin{center}
\begin{tabular}{|p{69pt}|p{34pt}|p{34pt}|p{34pt}|p{34pt}|p{34pt}|p{34pt}|p{34pt}|p{34pt}|p{34pt}|p{34pt}|p{34pt}|}
\hline
$\omega _L -\omega _M $ \par (10$^{13}$ rad/sec )&
$\bar {A}_{0}$ \par (\AA)&
$\bar {A}_{1}/\bar {A}_{0}$ &
$\bar {A}_{2}/\bar {A}_{0}$ &
$\bar {A}_{3}/\bar {A}_{0}$ &
$\bar {A}_{6}/\bar {A}_{0}$ &
$\bar {\xi}_{0}$ \par (\AA)&
$\bar {\xi}_{1}/\bar {\xi}_{0}$ &
$\bar {\xi}_{2}/\bar {\xi}_{0}$ &
$\bar {\xi}_{3}/\bar {\xi}_{0}$ &
$\bar {\xi}_{6}/\bar {\xi}_{0}$ &
$E$ \par (eV) \\
\hline
0.013&
0.116&
0.853&
0.534&
0.267&
0.050&
0.007&
0.571&
-0.143&
-0.286&
-0.0004&
0.204 \\
\hline
0.081&
0.142&
0.845&
0.514&
0.239&
0.023&
0.008&
0.750&
-0.250&
-0.375&
-0.0005&
0.255 \\
\hline
0.181&
0.180&
0.844&
0.522&
0.250&
0.034&
0.014&
0.571&
-0.143&
-0.357&
-0.0009&
0.366 \\
\hline
0.465&
0.314&
0.863&
0.564&
0.287&
0.022&
0.034&
0.618&
-0.059&
-0.235&
-0.003&
1.080 \\
\hline
0.672&
0.420&
0.905&
0.655&
0.365&
0.026&
0.048&
0.729&
0.146&
-0.354&
-0.006&
2.197 \\
\hline
0.742&
0.474&
0.945&
0.770&
0.508&
0.068&
0.040&
0.975&
0.525&
-0.175&
-0.011&
3.474 \\
\hline
0.779&
0.500&
0.966&
0.852&
0.640&
0.094&
0.030&
1.100&
1.000&
0.267&
-0.016&
4.596 \\
\hline
\end{tabular}
\label{tab1}
\end{center}
\end{table}

\end{widetext}

\subsection{Slow modulation of ILM}

Slow modulation of the ILM vibrational amplitude is observed in Fig. 3. The modulation 
is also seen as a satellite at the frequency $\omega_{LL}=5.6 \cdot 10^{13}$ 
rad/sec in the spectrum of vibrations of the bonds given in Fig. 5. The weak 
high-frequency satellite of the triple-peak in Fig. 5 results from the nonlinear 
four wave mixing  $2\omega_L -\omega_{LL}$. This effect has been studied in some detail 
for a monatomic 1-D lattice in Ref. [33] and has been identified with linear local modes 
(LLMs) associated with the lattice perturbation produced by ILM. 

To address the question whether the modulation observed for this 3-D system is connected 
with artificial size effects or LLMs we repeated the calculation for a cluster of twice 
larger size. No significant difference in the time dependence of vibrations of the central 
atoms have been observed. 
To perform an additional check on our interpretation of the origin of the 
modulation a second series of runs has been performed using as the initial conditions 
the long time displacement patterns of the 36 atoms (8 atoms in the central chain and 
7 atoms in every nearest 4 chains) more closely registered to the correct ILM shape. Now 
the starting amplitudes of the central particles are determined by averaging over 
a modulation period. The time dependence of the vibration of the central ($n$=0) and 
third ($n$=3) bonds calculated in this way are presented in Fig. 6. Since the starting 
shapes more closely resemble a pure ILM eigenvector the only significant difference from 
the results shown in Fig. 3 is the reduction of the amplitude of modulation. The periods 
of the ILM vibration and of the modulation remain unchanged, precisely what is expected 
for an LLM of smaller amplitude. This removes the possibility that the observed modulation 
is associated with the reflection of phonon wave packets from the cluster boundary.

This conclusion also agrees with our calculation of vibrations of atoms near the border. 
The amplitudes of these border vibrations always remain less or of the order of 
$5\cdot10^{-3}$ \AA, i.e. they are much smaller than the amplitude of modulation of 
vibrations of the central atoms. 

All of these signatures allow us to assign the observed 
modulation to the linear local mode produced by the ILM. Indeed, as expected for a LLM, 
its frequency increases with increasing amplitude and frequency of the ILM.  An 
enlargement of the modulation for the side bonds shows, in agreement with Ref. [33], that 
the maximum amplitude of the LLM is situated in the periphery of the ILM. An interesting 
property of the mode, which causes the modulation, is the change of the sign of the 
frequency difference $\omega_{L}-\omega_{LL}$  with decreasing localization of the ILM. 
For the ILM in the first line of the Table I $\omega_{L}-\omega_{LL}$ is negative 
($-0.14\cdot 10^{13}$ rad/sec), while for ILMs in all other lines it is positive 
(0.17, 0.24, 0.42, 0.45, 0.38 and 0.35, respectively; all in $10^{13}$ rad/sec units). 
This change of sign is also expected for the LLM: it results from the 
different dependence of the ILM and LLM on the even anharmonicities [33].

In order to verify the conclusion about the existence of ILMs in Ni at elevated 
temperatures we repeated the calculations presented in Fig. 6 for the lattice constant 
corresponding to 800 K. We have found that the ILM also exists at this temperature but 
with slightly enlarged ($\sim 3\%$) amplitude and reduced frequency
$\omega_{L}=5.67\cdot10^{13}$ rad/sec. Thermal fluctuations characteristic for non-zero 
temperature in the initial state have been ignored; however thermal-like fluctuations 
appeared in our MD simulations in the latter stages of the time-evolution of the system 
due to shaking-off of the phonons. These fluctuations do not significantly affect the ILM.
\vspace{0.4cm}

\begin{figure}[th]
\includegraphics[angle=0,width=.48\textwidth]{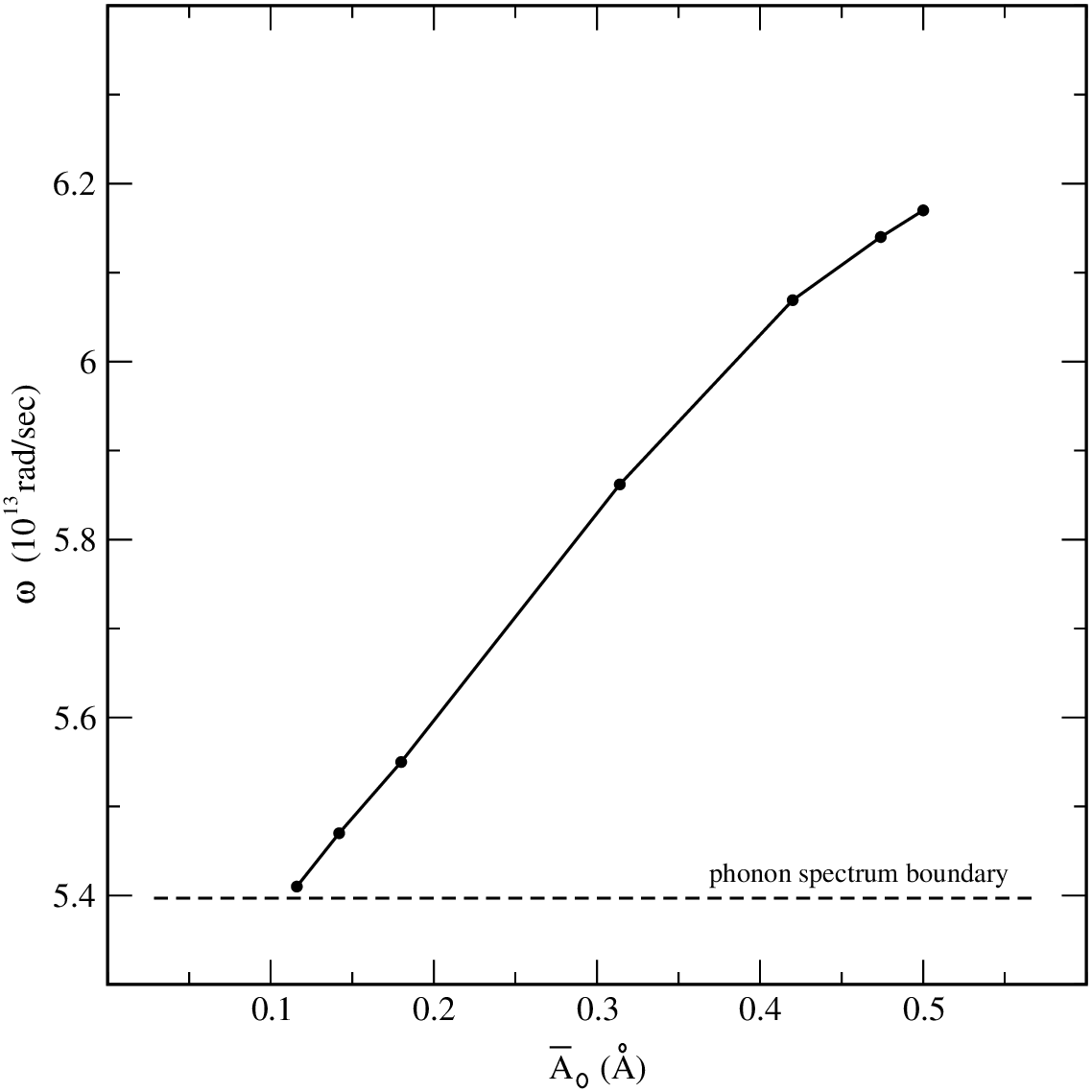}\hspace*{0em}
\caption{The dependence of frequency $\omega_L$ of the even ILM in
Ni on the amplitude of vibrations of the central bond.}
\end{figure}

\begin{figure}[th]
\includegraphics[angle=0,width=.48\textwidth]{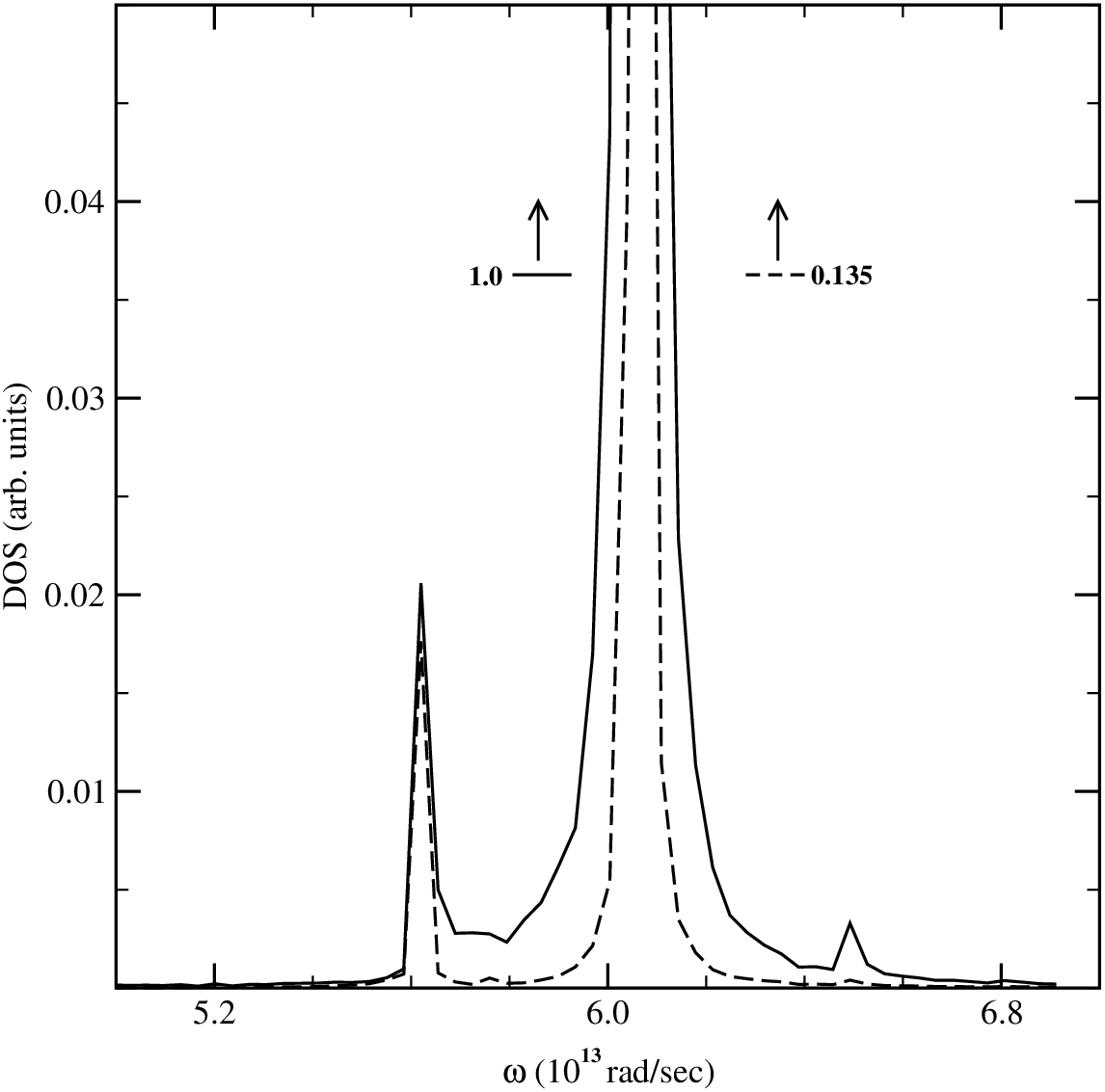}\hspace*{0em}
\caption{ILM with its satellite LLM ($\omega_L=6.07 \cdot 10^{13}$ rad/sec).
Shown are the Fourier transforms of vibrations $\bar {u}_0 (t)$ (solid line)
and $\bar {u}_3 (t)$ (dashed line) given in Fig. 3. }
\end{figure}

\begin{figure}[th]
\includegraphics[angle=0,width=.48\textwidth]{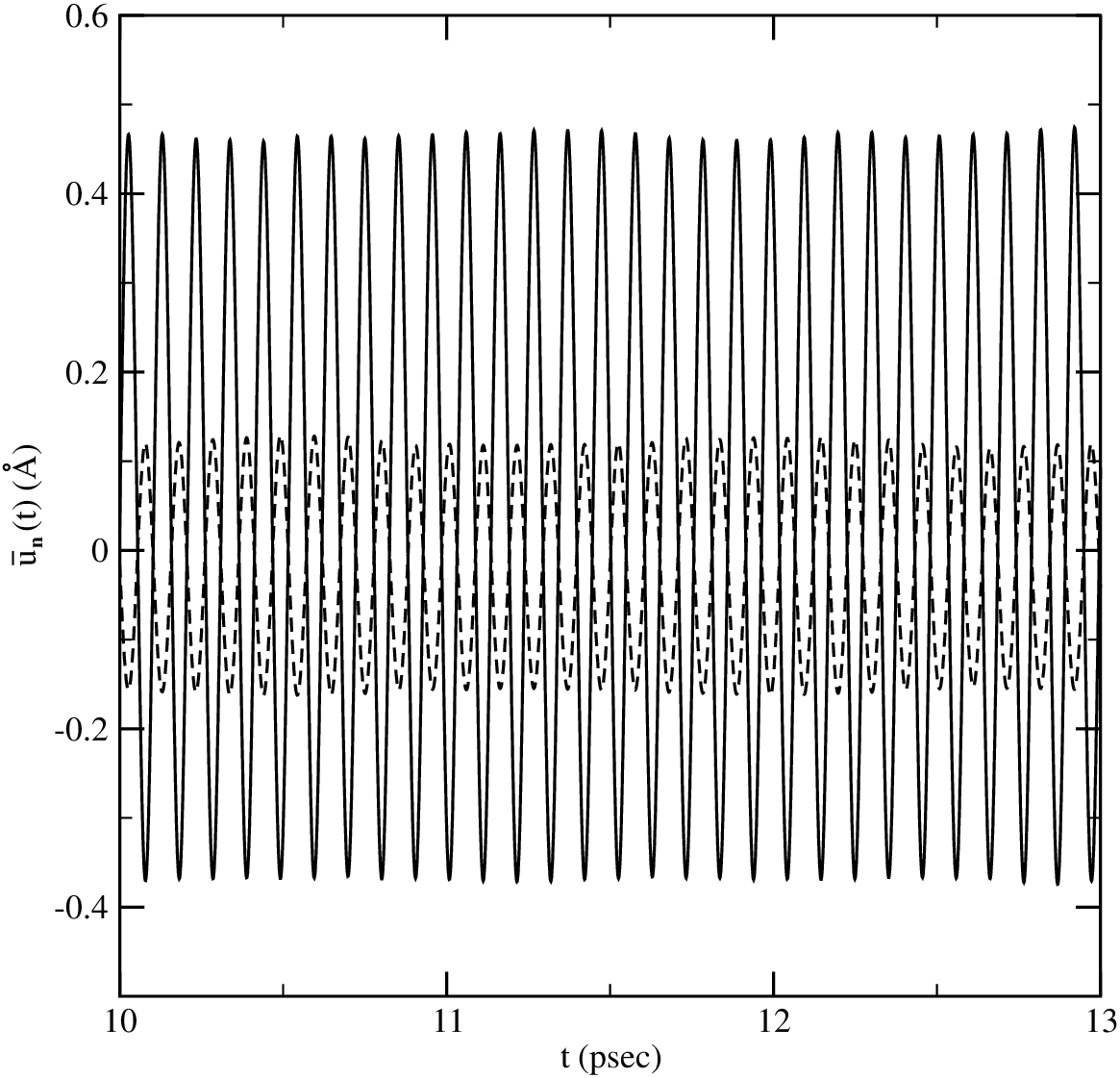}\hspace*{0em}
\caption{The same time dependence $\bar {u}_n (t)=r_n(t)-r_0$ of the vibration of the 
central ($n$=0, solid line) and third ($n$=3, dashed line) bonds in Ni at long times
as shown in Fig. 3. The initial shifts of the 36 central atoms (8 atoms in the
central chain and 7 atoms in every nearest 4 chains) more closely described by the 
correct ILM shape. The starting values of the shifts are obtained by averaging over
a modulation period in Fig. 3. Since the ILM shape is more closely approximated the
modulation amplitude is greatly reduced as compared to Fig. 3. }
\end{figure}

\subsection{ILMs in niobium}

In contrast with nickel, niobium is isotopically pure and may provide a cleaner 
experimental signature for the observation of intrinsic localization. Using an analogous 
algorithm we also performed molecular dynamic simulations of ILMs in Nb. 
The peculiarity of this metal is the rather slow screening of the interactions
with increasing distance between the ions. For a correct
description of the phonon DOS one needs to take into account the elastic
forces between at least six nearest neighboring atoms [34].

The existing EAM theories of Nb do not describe sufficiently well atomic forces for
so large a spatial interval. Therefore we did not use them in a description of the
linear dynamics. Instead, for this purpose the force constants for 6 nearest neighbors
given in Ref. [34] were used. Thus, for Nb we did not use the single
pair-potential for MD simulations (as we did it for Ni). Instead we used 6
different potentials (more precisely, we used 6 different forces) for every
6 nearest atom pairs.

The EAM potential of Ref. [35] was used only to find the anharmonic forces, namely,
the 3 nonlinear forces for 3 nearest atoms. Corresponding to this EAM pair potential
$V(r)$ as well as its approximation by the fourth-order polynomial are presented
in Fig. 7 (corresponding parameters equal 
$K_2 \approx 1.5 \, \textrm{eV} / {\textrm{\AA}}^2$, 
$K_3 \approx -6.2 \, \textrm{eV} / {\textrm{\AA}}^3$, 
$K_4 \approx 65.6 \, \textrm{eV} / {\textrm{\AA}}^4$;
the root-mean-square deviation for the polynomial approximation in the interval between
2.82 and 2.97 \AA \, is  10$^{-5}$ eV). This $V(r)$ equals to the first term of EAM given
by Eq. (3) in Ref. [35] plus the linear at $r_0$ part  of the second term in this
equation ($r_0$ is  the equilibrium distance of the nearest atoms at given temperature).
The anharmonic forces were found as follows: for each of 3 nearest atom pairs we deleted
from the given in Fig. 7 potential the linear and quadratic terms at equilibrium distance.
Then we added to each of these nonlinear forces the corresponding linear force given by
Ref. [34]. 

The same source was used to obtain the remaining three linear forces
between the atom pairs at larger distances. The justification for this
approximation is that the relative change in distance of the fourth, fifth and
sixth atom-pairs for the ILMs with the maximum amplitude $\sim $ 0.1 \AA \, is
few percent. For such small relative shifts and large distances the
anharmonic forces are negligibly small. Therefore for a description of ILMs
with small amplitude, one may use the known elastic forces for six nearest neighbors [34]
and to take into account the anharmonic forces only for three nearest neighbors.

\begin{figure}[th]
\includegraphics[angle=0,width=.48\textwidth]{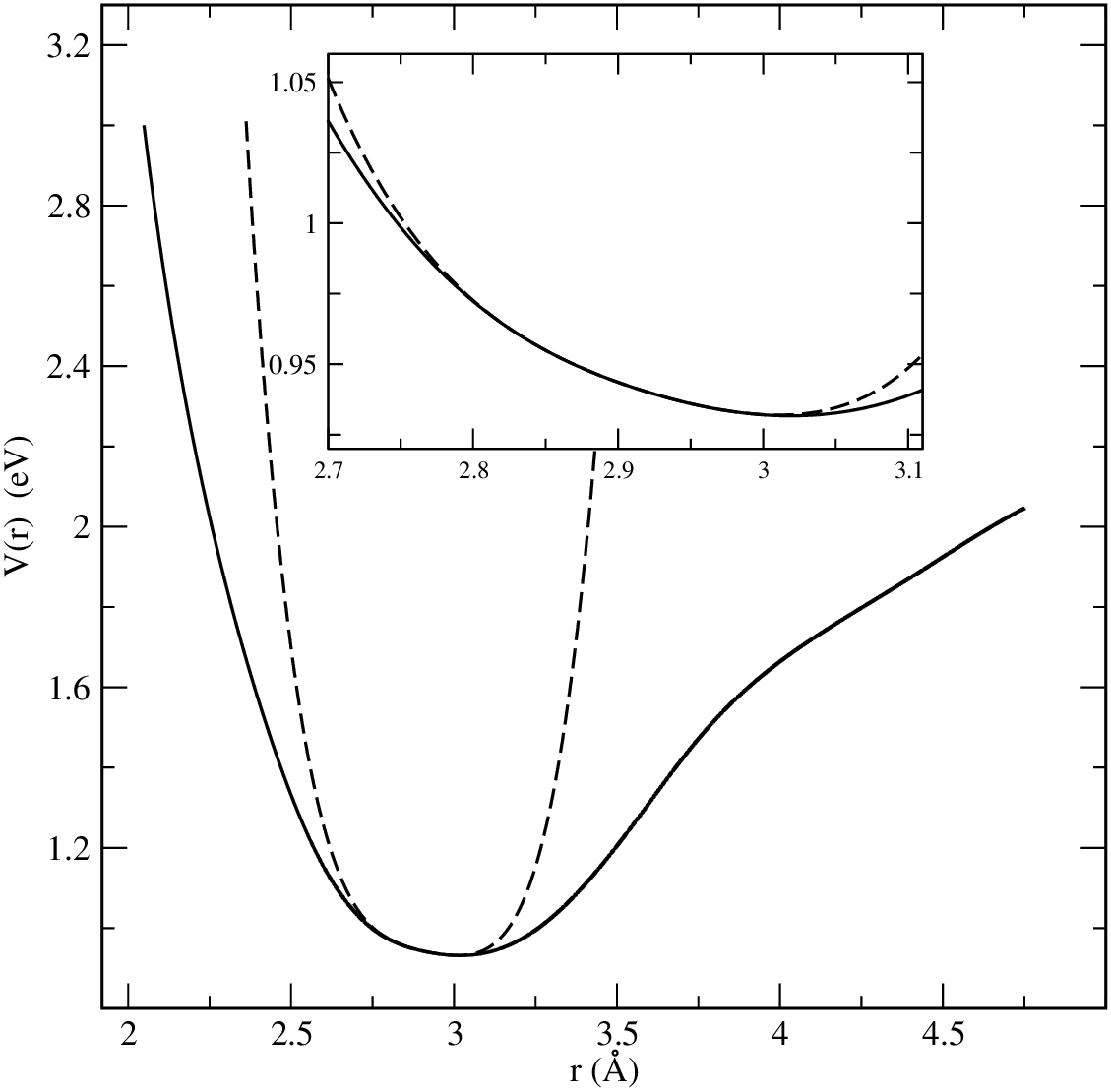}\hspace*{0em}
\caption{The pair potential $V(r)$ of Nb (solid line) and its approximation by the
fourth-order polynomial (dashed line). Insert shows an expanded view.}
\end{figure}

Using the algorithm described above, we performed MD simulations of ILMs
in  Nb for room temperature 293 K  (equilibrium first neighbor distance 
$r_0$=2.86 \AA, longitudinal velocity of sound $ v_l=$5380 m/sec) and  high
temperature 1773 K ($r_0$= 2.89 \AA, $v_l =$ 5073 m/sec). For both temperatures
$\tilde {\kappa} > 1$, i.e.  the derived condition of an ILM with frequency above the
phonon spectrum is fulfilled. The calculations at room temperature were made for
a cluster containing  2 x 40 x 40 x 40 moving atoms (altogether 128000 atoms).
The calculations at high temperature were made for a cluster elongated in the [111]
direction with $C_{3h} $ symmetry (a hexagonal prism) containing 18760 moving atoms.
To minimize the calculation time, the positions and velocities of 1/6 of the atoms,
situated in one of the six identical segments of the prism
were calculated at every time step; the positions and velocities of all other atoms
were found from the symmetry conditions.

For both sets of lattice parameters we have found even ILMs in the
[111] direction of vibrations of the central atoms with frequencies above the
top of the phonon spectrum. These ILMs are fully stable: no decay of their
amplitude was observed over the last 500 periods of vibrations. A small
periodic modulation of the amplitude, analogous to Ni was also observed,
presumably caused by the appearance of a linear local mode. The spectra of
ILMs for two different amplitudes of vibrations of the central bond are presented in
Fig. 8 and 9 (together with the phonon spectrum). As expected the frequency of the ILM
increases with amplitude.

\begin{figure}[th]
\includegraphics[angle=0,width=.48\textwidth]{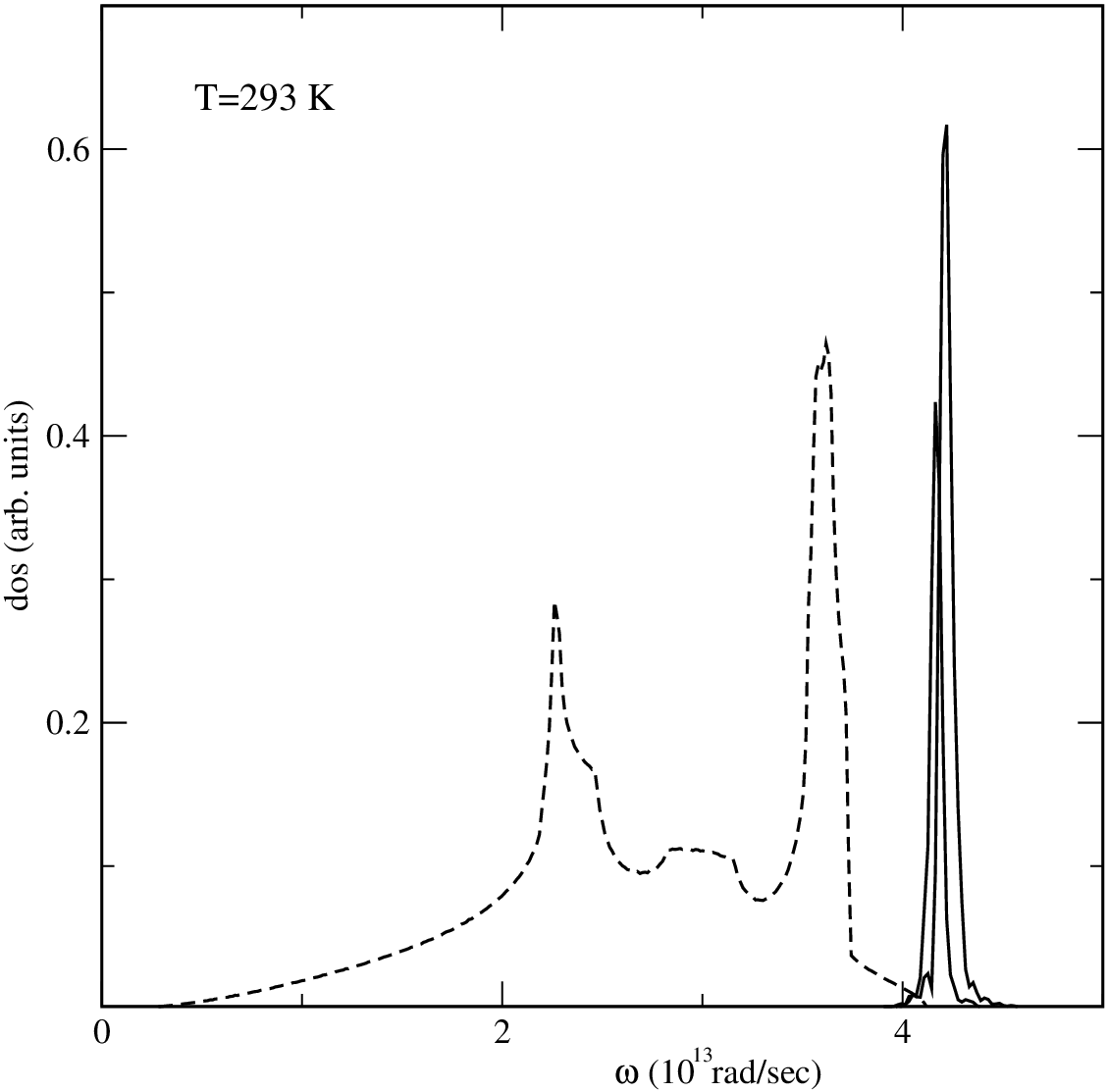}\hspace*{0em}
\caption{Phonon DOS and spectra of even
ILMs in Nb at room temperature 293 K for two amplitudes (0.25 \AA \, and 0.3 \AA)
of vibrations of the central bond. Solid lines: ILMs, dashed line: phonon spectrum of Nb.}
\end{figure}

\begin{figure}[th]
\vspace{0.8cm}\includegraphics[angle=0,width=.48\textwidth]{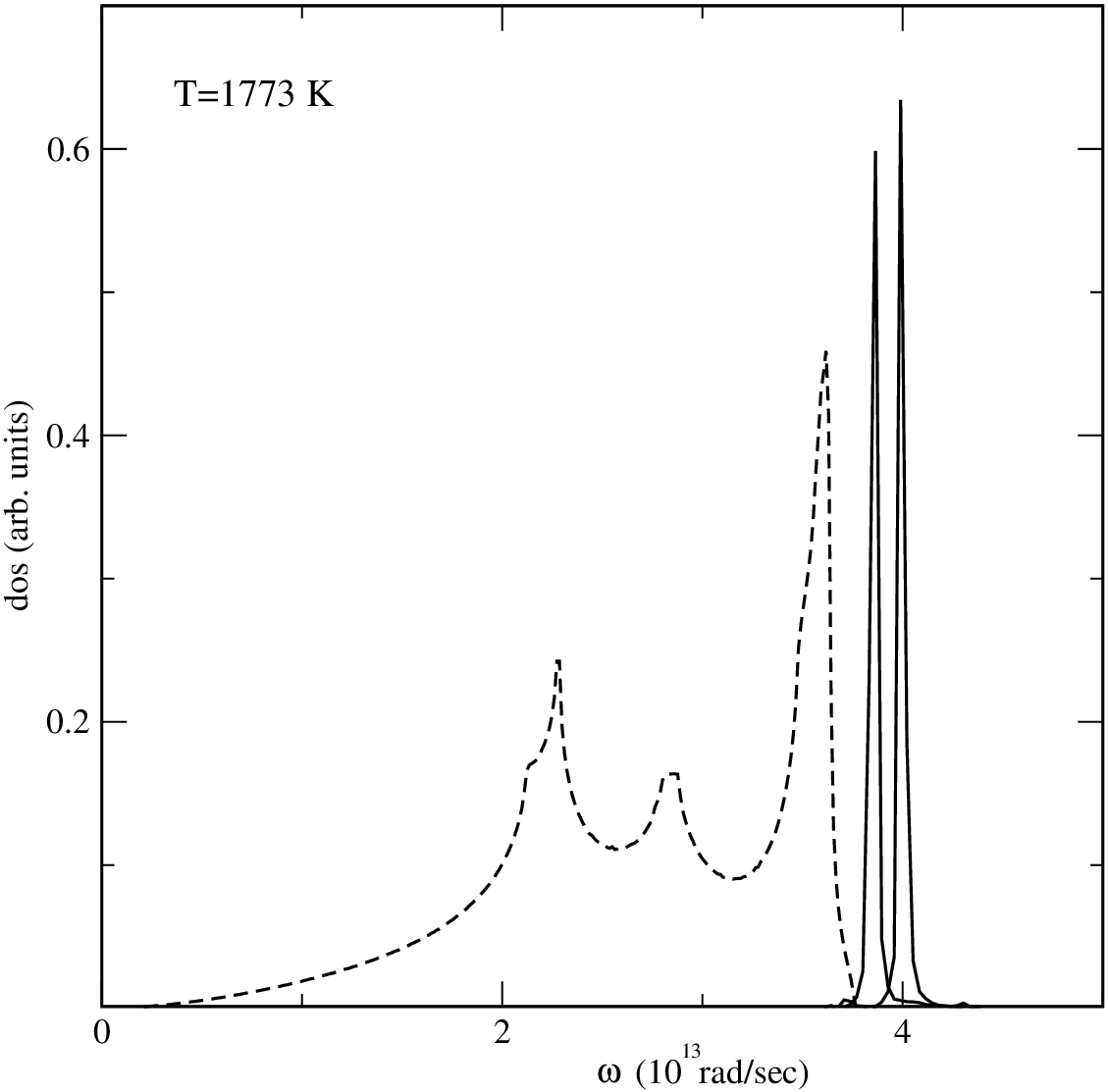}\hspace*{0em}
\caption{Phonon DOS and spectra of even ILMs in Nb at a high temperature (1773 K)
for two amplitudes (0.085 \AA \, and 0.12 \AA) of vibrations of the central bond.
Solid lines: ILMs, dashed line: phonon spectrum of Nb.}
\end{figure}

To check the specificity of the above procedure we also performed calculations of ILMs 
in Nb taking into account the full potential given in Ref. [35] (including the 
volume-dependent second term in Eq. (10)). At high temperatures stable ILMs do exist 
for this model. In addition, we found that the formation of ILMs in Nb is favored by 
the expansion of the Nb lattice with increasing temperature.

On the other hand, based on the relations presented here, we have found that at least 
in some other metals (e.g. in Al and Cu) ILMs of the type described here should not exist. 
Indeed the values of the parameter $\tilde {\kappa}$ in these metals at room temperature 
are found to be 0.3 (Al) and 0.38 (Cu) (using the potentials given in Ref. [32]). At 800 K 
these parameters correspondingly equal 0.1 and 0.42. All these values are much smaller 
than the border value 1, which makes it unlikely that ILMs of the type described here can 
appear in these metals.

\section{Conclusion}
To sum up we performed an analytic and numerical study of nonlinear dynamics of Ni and Nb
and have found that intrinsic localized modes may exist in these metals with frequencies
above the top of the phonon bands. The physical reason for this is the relatively large 
value of even anharmonicities as compared to odd ones produced by the free electron 
screening of the atomic interactions. As a result, in Ni and Nb (and, presumably in some
other metals), the ion-ion attractive force, at intermediate distances is enhanced
resulting in the amplification of even anharmonicities for the two-body potentials.
This effect counteracts the underlying softening associated with the bare potentials
with moderate increase of vibrational amplitudes to permit the existence of ILMs
above the top of the phonon spectrum. In our MD simulations of nonlinear dynamics of Ni 
and Nb we have clearly observed ILM of this type. In addition we also observed
the linear local modes associated with an ILM; modes of this-type have been recently
predicted and observed numerically for chains in Ref. [33]. According to our calculations 
an intrinsic localized mode in Ni or in Nb may have a rather small amplitude and hence 
a small energy of formation. We expect that in these metals ILMs may be observed as 
a high frequency features in the phonon spectrum at high temperatures. Finally we note 
that the precision of the existing EAM models of Nb is rather low [35, 36]; however, 
the EAM models of Ni are usually considered to be quite precise [31]. Therefore 
the conclusions presented here about ILMs in Ni are expected to be more reliable 
than those about ILMs in Nb. 

\section{Acknowledgments}
The research was supported by ETF grants no. 7741 and 7906 and by the European Union 
through the European Regional Development Fund (project TK114). AJS was supported by 
NSF-DMR-0906491.


\vspace{5mm}

\begin{thebibliography}{99}

\bibitem{1}
B. R. Henry, J. Phys. Chem. \textbf{80}, 2160 (1976).

\bibitem{2}
M. L. Sage and J. Jortner, Adv. Chem. Phys. \textbf{47}, 293 (1981).

\bibitem{3}
A. A. Ovchinnikov and N. S. Erihjman, Sov. Phys. Usp. \textbf{25}, 738
(1982).

\bibitem{4}
B. R. Henry and H. G. Kjaergaard, Can. J. Chem. \textbf{80}, 1635
(2002).

\bibitem{5}
A. M. Kosevich and A. S. Kovalev, Zh. Eksp. Teor. Fiz. {\bf 67},
1793 (1974) [Sov. JETP {\bf 40}, 891 (1974)].

\bibitem{6}
A. S. Dolgov, Fiz. Tverd. Tela (Leningrad) {\bf 28}, 1641 (1986)
[Sov. Phys. Solid State {\bf 28}, 907 (1986)].

\bibitem{7}
A. J. Sievers and S. Takeno, Phys. Rev. Lett. \textbf{61}, 970 (1988).

\bibitem{8}
J. B. Page, Phys. Rev. B \textbf{41}, 7835 (1990).

\bibitem{9}
R. S. MacKay and S. Aubry, Nonlinearity \textbf{7}, 1623 (1994).

\bibitem{10}
S. A. Kiselev and V. I. Rupasov, Phys. Lett. A \textbf{148}, 355
(1990).

\bibitem{11}
S. R. Bickham, A. J. Sievers, and S. Takeno, Phys. Rev. B \textbf{45},
10344 (1992).

\bibitem{12}
K. W. Sandusky, J. B. Page, and K. E. Schmidt, Phys. Rev. B
\textbf{46}, 6161 (1992).

\bibitem{13}
A. J. Sievers and J. B. Page, in \textit{Dynamical Properties of Solids: 
Phonon Physics The Cutting Edge}, edited by G.K. Norton and A.A.
Maradudin (North Holland, Amsterdam, 1995), Vol. VII, p. 137.

\bibitem{14}
S. Flach and C. R. Willis, Phys. Repts. \textbf{295}, 182 (1998).

\bibitem{15}
R. Lai and A. J. Sievers, Phys. Repts. \textbf{314}, 147 (1999).

\bibitem{16}
D. K. Campbell, S. Flach, and Y. S. Kivshar, Physics Today \textbf{57}(1),
43 (2004).

\bibitem{17}
M. Sato, B. E. Hubbard, and A. J. Sievers, Rev. Mod. Phys. \textbf{78},
137 (2006).

\bibitem{18}
S. Flach and A. Gorbach, Phys. Repts. \textbf{467}, 1 (2008).

\bibitem{19}
S. A. Kiselev, S. R. Bickham, and A. J. Sievers, Phys. Rev. B
\textbf{48}, 13508 (1993).

\bibitem{20}
S. A. Kiselev and A. J. Sievers, Phys. Rev. B \textbf{55}, 5755 (1997).

\bibitem{21}
V. Hizhnyakov, D. Nevedrov, and A. J. Sievers, Physics B
\textbf{316-317}, 132 (2002).

\bibitem{22}
L. Z. Khadeeva and S. V. Dmitriev, Phys. Rev. B \textbf{81}, 214306
(2010).

\bibitem{23}
M. E. Manley, M. Yethiraj, H. Sinn, H. M. Volz, A. Alatas, J. C. Lashley, W. L.
Hults, G. H. Lander, and J. L. Smith, Phys. Rev. Lett. \textbf{96}, 125501
(2006).

\bibitem{24}
M. S. Daw and M. I. Baskes, Phys. Rev. B \textbf{29}, 6443 (1984).

\bibitem{25}
M. S. Daw and M. I. Baskes, Phys. Rev. Lett. \textbf{50}, 1285 (1983).

\bibitem{26}
W. A. Harrison, \textit{Solid State Theory} (McGraw Hill, New York, 1970).

\bibitem{27}
V. Hizhnyakov, A. Shelkan, and M. Klopov, Phys. Lett. A \textbf{357},
393 (2006).

\bibitem{28}
A. Shelkan, V. Hizhnyakov, and M. Klopov, Phys. Rev. B \textbf{75},
134304 (2007).

\bibitem{29}
B. Sanchez-Rey, G. James, J. Cuevas, and J. F. R. Archilla, Phys. Rev. B \textbf{70}, 
014301 (2004).

\bibitem{30}
S. R. Bickham, S. A. Kiselev and A. J. Sievers, Phys. Rev. B \textbf{47}, 14206 (1993).

\bibitem{31}
Y. Mishin, D. Farkas, M. J. Mehl, and D. A. Papaconstantopoulos, Phys.
Rev. B \textbf{59}, 3393 (1999).

\bibitem{32}
http://cst-www.nrl.navy.mil/bind/eam.


\bibitem{33}
V. Hizhnyakov, A. Shelkan, M. Klopov, S. A. Kiselev, and A. J. Sievers,
Phys. Rev. B \textbf{73}, 224302 (2006).

\bibitem{34}
F. Guthoff, B. Hennion, C. Herzig, W. Petry, H. R. Schober, and J.
Trampenau, J. Phys. Cond. Mat. \textbf{6}, 6211 (1994).


\bibitem{35}
M. R. Fellinger, H. Park, and J. W. Wilkins, Phys. Rev. B \textbf{81},
144119 (2010).

\bibitem{36}
A. M. Guellil and J. B. Adams, J. Mater. Res. 7, 639 (1992).
\end{thebibliography}
\end{document}